\documentclass[aps,prl,twocolumn,showpacs,groupedaddress]{revtex4}

\usepackage{graphicx}

\newcommand{\etal}{{\it et~al.}}

\begin{document}

\preprint{}

\title{Carrier Freeze-Out Induced Metal-Insulator Transition
in Oxygen Deficient SrTiO$_3$ Thin Films}

\author{Z. Q. Liu$^{1,2}$}

\author{D. P. Leusink$^{1,2,4}$}

\author{X. Wang$^{1,2}$}

\author{W. M. L\"{u}$^{1,3}$}

\author{K. Gopinadhan$^{1,3}$}

\author{A. Annadi$^{1,2}$}

\author{Y. L. Zhao$^{1,2}$}

\author{X. H. Huang$^{1,3}$}

\author{S. W. Zeng$^{1,2}$}

\author{Z. Huang$^{1}$}

\author{A. Srivastava$^{1,2}$}

\author{S. Dhar$^{1,3}$}

\author{T. Venkatesan$^{1,2,3}$}

\author{Ariando$^{1,2}$}

\altaffiliation[Email: ]{ariando@nus.edu.sg}

\affiliation{$^1$NUSNNI-Nanocore, $^2$Department of Physics, $^3$Department of Electrical and Computer Engineering, National University of Singapore, Singapore}

\affiliation{$^4$Faculty of Science and Technology and MESA$^+$ Institute for Nanotechnology, University of Twente, 7500 AE Enschede, The Netherlands}

\date{\today}

\begin{abstract}
We report optical, electrical and magneto transport properties of a high quality oxygen deficient SrTiO$_3$ (STO) thin film fabricated by pulsed laser deposition technique. The oxygen vacancy distribution in the thin film is expected to be uniform. By comparing its electrical properties to those of oxygen deficient bulk STO, it was found that the oxygen vacancies in bulk STO is far from uniform over the whole material. The metal-insulator transition (MIT) observed in the oxygen deficient STO film was found to be induced by the carrier freeze-out effect. The low temperature frozen state can be re-excited by Joule heating, electric and intriguingly magnetic field.
\end{abstract}



\maketitle

Like Silicon in the semiconductor technology, SrTiO$_3$ (STO) is the
most used substrate in oxide electronics because of its large
dielectric constant [1], the close lattice match to a wide range of
other perovskite oxides and its excellent thermal and chemical
stabilities. Recently, two-dimensional electron gas [2,3] and
electronic phase separation [4] have been demonstrated to emerge at
the bare STO surface. Understanding the electronic and magnetic
properties of STO under different oxidation states and external
excitations is therefore very crucial to reveal the origin of these
emerging phenomena and to use STO in devices. STO is a typical
nonpolar band insulator with an indirect band gap of $\sim3.27$ eV
[5], but oxygen deficient STO can show a metallic phase with a
flexible tunability in electrical conductivity depending on the
concentration of oxygen vacancies. More attractively, oxygen
deficient STO is the first oxide discovered to be superconductive
[6] with the $T_c$ between 0.1 and 0.6 K [7]. Shubnikov-de Haas
(SdH) oscillations [8] are also observed under usual laboratorial
magnetic fields due to the high mobility [9] of free electrons. As a
result of the large effective mass of the electrons at the bottom of
the conduction band, oxygen deficient STO possesses a large Seebeck
coefficient of $\sim890$ $\mu$V/K at room temperature (RT) [10],
thus being a focus point in solid state thermoelectric too.

Although so many interesting phenomena in oxygen deficient STO, the
inhomogeneity of oxygen vacancies obtained by reducing bulk single
crystals in vacuum and at high temperature is always an issue
[6-11]; that is because the diffusion process of oxygen ions is
usually proportional to the square root of time and the diffusion
coefficient of oxygen ions in STO is small, namely $\sim10^{-10}$
cm$^2$/s at 950$^\circ$C [12]. For example, taking 1 hour as the
annealing time at 950$^\circ$C, the diffusion length $l=\sqrt{Dt}$
of oxygen ions in single crystal STO will be $\sim6$ $\mu$m, where
$D$ is the diffusion coefficient and $t$ is the annealing time.
Tufte and Chapman [9] found that the reduced STO samples begin to
reoxidize from a very low temperature of $\sim$500 K. Frederikse
{\etal} [8] observed SDH oscillations in reduced STO samples, which
theoretically and also experimentally prefer to appear in a system
close to a high mobility two-dimensional electron system [13]. These
observations suggest that the oxygen vacancies may mostly exist near
the surface rather than uniformly over the whole bulk material.
Therefore, the values of some physical quantities [6-11] like
resistivity and carrier density, derived from the thickness of the
whole bulk sample, which was used to characterize the dimension of
the conducting area, would be of dubious validity. On the other
hand, the intrinsic properties of the material are closely related
to these physical parameters. For example, the superconducting $T_c$
of oxygen deficient STO apparently depends on the carrier density
[7,11]. Additionally, the inhomogeneities could generate significant
influence on electrical and galvanomagnetic measurements, and can
even give rise to intriguing quantum effect-linear magnetic field
dependence of the transverse magnetoresistance (MR) [14,15].

\begin{figure}
\includegraphics[width=3.4in]{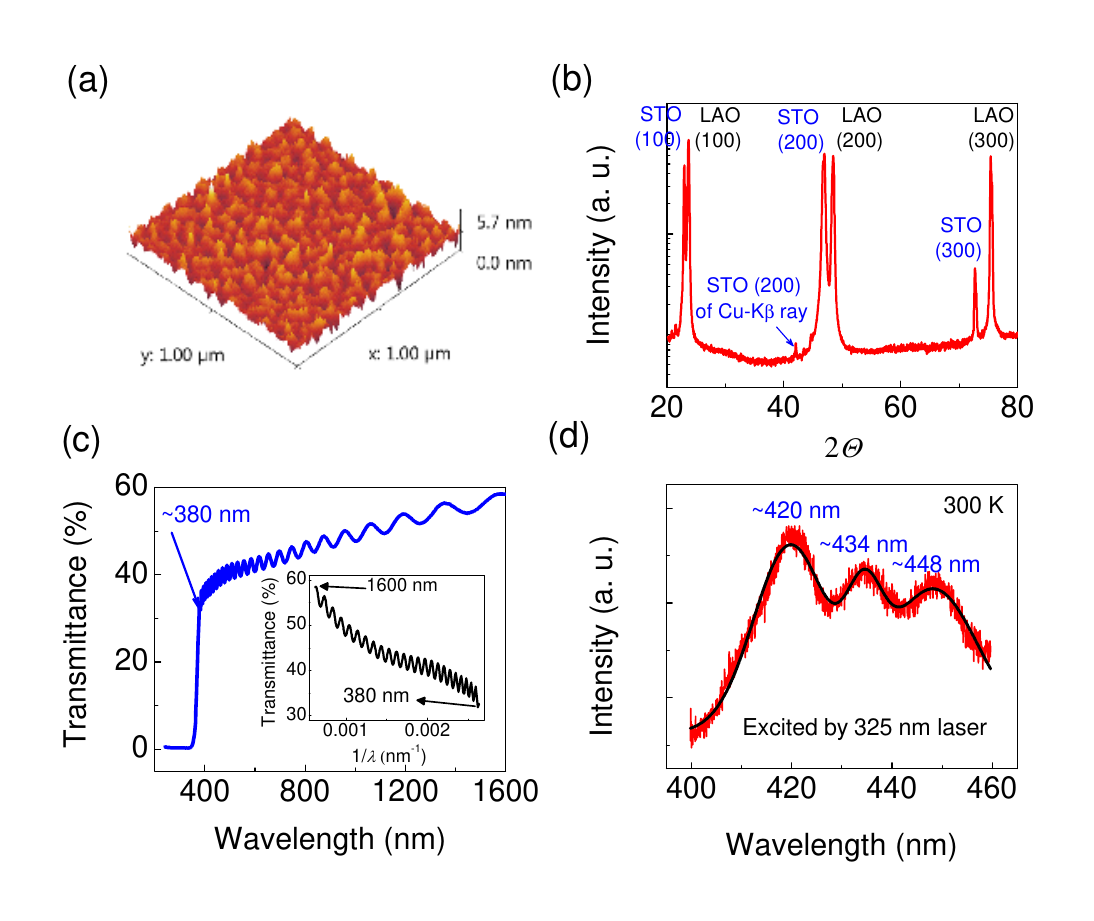}
\caption{\label{fig1} (a) 3D atomic force microscope image of an
$1\times 1$ $\mu$m$^2$ area of the STO film. (b) X-ray diffraction
of as-grown STO film on LAO substrate. (c) Room temperature
ultraviolet-visible-infrared spectroscopy of the oxygen deficient
STO film (obtained by annealing in $\sim 1\times 10^{-7}$ Torr
vacuum at 950$^\circ$C for 1 hour) from 240 to 1600 nm. (Inset) The
transmittance data plotted versus the reciprocal of the wavelength
from 380 to 1600 nm. (d) Room temperature photoluminescence
spectroscopy of the oxygen deficient STO film between 400 and 460
nm.}
\end{figure}

\begin{figure}
\includegraphics[width=3.4in]{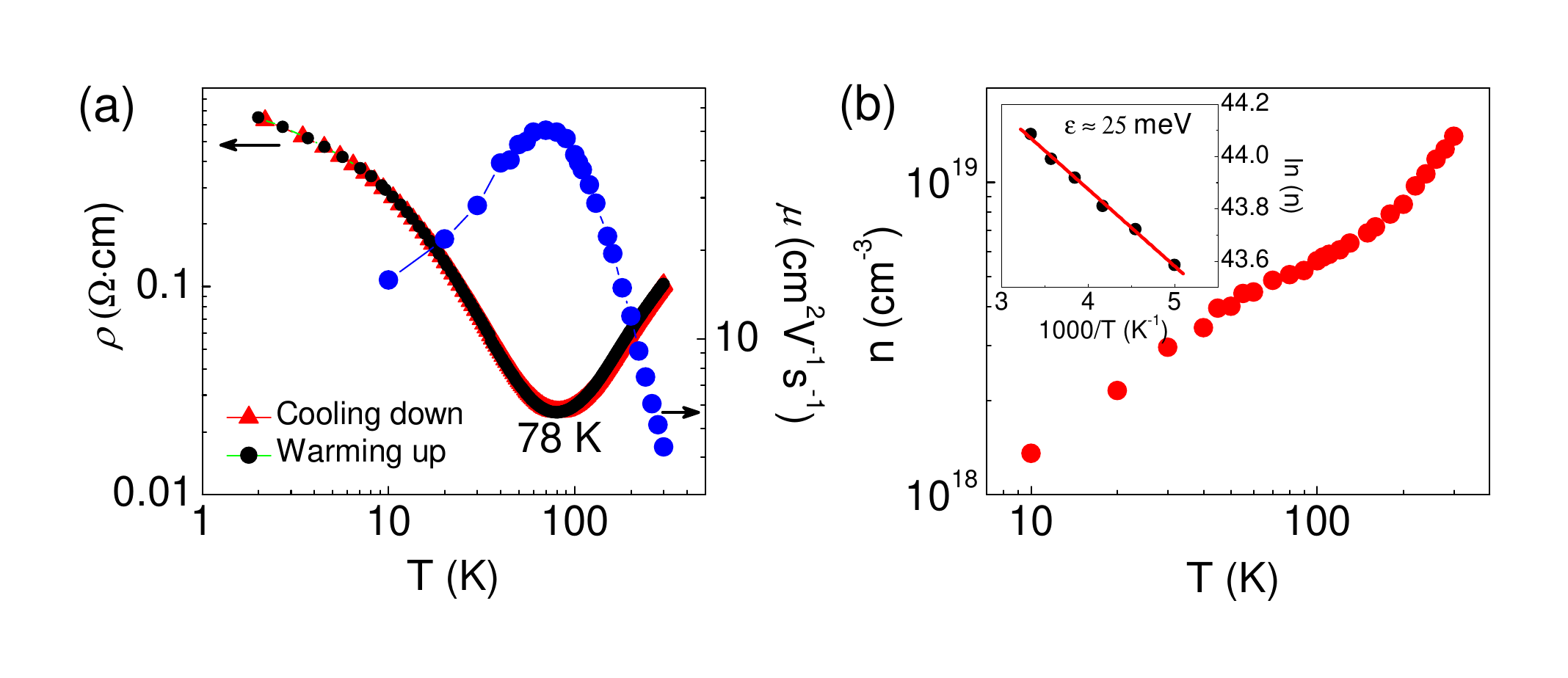}
\caption{\label{fig2} (a) Temperature dependences of the resistivity
$(\rho-T)$ during different measurement processes and its carrier
mobility over the temperature range $300-2$ K, and (b) the
corresponding carrier density $n$. The Arrhenius plot of $\ln(n)$
for the temperature range $300-200$ K and the linear fitting are
drawn in the inset of (b).}
\end{figure}

Here we report the metal-insulator transition (MIT) observed in
oxygen deficient STO film, in which the oxygen vacancies are
expected to uniformly distribute. As a highly interesting subject in
condensed matter physics, MIT has various intriguing mechanisms
[16]. However the MIT observed here is ascribed to the deionization
effect of oxygen vacancies with decreasing temperature, which serve
as doubly charged donor centers to make STO metallic at high
temperatures. Both the resistivity and carrier density are
significantly different from the ones of the bulk samples with the
carrier freeze-out  phenomenon [9], {\it{i.e.}}, several ten times
smaller and larger respectively. The frozen non-metallic state can
be re-excited by electric field and Joule heating. Surprisingly, it
was found that the low temperature carrier freeze-out can also be
suppressed by large magnetic fields, leading to negative MR.

In this work, pulsed laser deposition technique was used to
fabricate a STO film from a single crystal STO target on a
(100)-oriented LaAlO$_3$ (LAO) single crystal substrate (both sides
polished) under $6\times10^{-3}$ Torr O$_2$ at 800$^\circ$C. During
deposition, the fluence of laser energy was kept at 4 J/cm$^2$ and
the repetition rate of laser 4 Hz. To avoid the influence of the
possible interface effect [13] and to keep the properties of the
film akin to bulk STO, a film with a thickness of $\sim$2.6 $\mu$m
was deposited.

\begin{figure}
\includegraphics[width=1.4in]{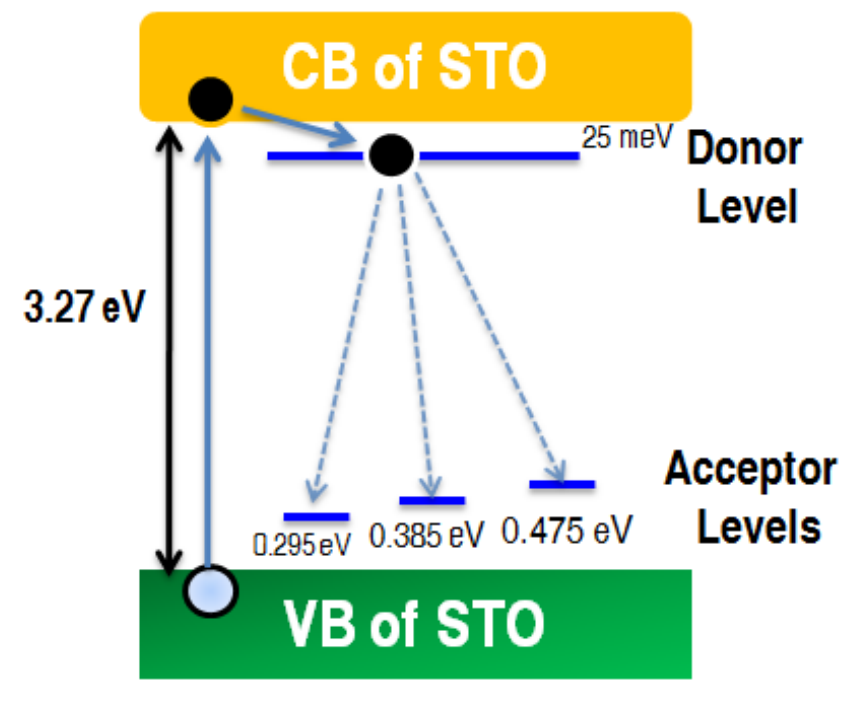}
\caption{\label{fig3}Band diagram of the oxygen deficient STO film and also the possible light emission mechanism. The energy intervals are not drawn to scale.}
\end{figure}

The 3D atomic force microscope image of the topography of the
as-deposited STO film on the LAO substrate [Fig. 1(a)] shows the
variation in the $z$-dimension which is less than 6 nm. The root
mean square value of the surface roughness in this 1 $\mu$m $\times$
1 $\mu$m area is only $\sim$0.541 nm, revealing a quite flat surface
considering the film thickness of $\sim$2.6 $\mu$m determined by
surface profile measuring system. The x-ray diffraction
(Cu-K$_\alpha1$ ray) pattern of the STO film is shown in Fig. 1(b).
The adjacent double diffraction peaks of the film and the substrate
for each order indicate the typical characteristic of epitaxial
growth. LAO is an excellent substrate to study oxygen deficient
films because it is fairly difficult to create oxygen vacancies
inside. Although it is a polar material, the bulk LAO will
experience surface reconstruction and thus its surface would be
nonpolar. This was con?rmed by our experiments, where STO
layer-by-layer grown on fully LaO-terminated LAO was highly
insulating, indicating no polar discontinuity at the interface.

The deposited STO film was reduced by annealing the sample at
950$^\circ$C and $\sim1\times10^{-7}$ Torr vacuum for 1 hour. The
transmittance spectrum [Fig. 1(c)] of the oxygen deficient STO film
was measured using ultraviolet-visible (UV) spectroscopy from 240 to
1600 nm. The absorption edge is $\sim$380 nm, well corresponding to
the band gap of STO. This suggests that after vacuum annealing the
lattice structure of the STO film is still well standing. The
spectrum displays an interference pattern with relatively high
transmittance above 380 nm, which pertains to a scenario where the
thick and highly smooth STO film serves as a Fabry-Perot
interferometer. So the transmittance will show peaks when the
wavelength $\lambda$ of the normal incident optical wave meets the
following condition:
\begin{equation}
2n(\lambda)d = k\lambda
\end{equation}
where $n(\lambda)$ is the refractive index as a function of
wavelength due to the strong dispersion [17], $d$ is the film
thickness and $k$ is the order of the fringe. Equivalently, the
transmittance will periodically oscillate with $n(\lambda)/\lambda$
at a period of $1/2d$. To simply estimate the film thickness from
the oscillations, take an intermediate refractive index 2.0516 [18],
to keep $n$ as a constant. Thus if the transmittance data of the
wavelength from 380 to 1600 nm are plotted versus $1/\lambda$ as
shown in the inset of Fig. 1(c), the average period of the
oscillations in $1/\lambda$ will be $1/2nd$. The fitted average
period is $9.66384\times 10^{-5}$ nm$^{-1}$ and therefore the
derived thickness is 2524 nm. The value is consistent with the
directly measured one but slightly smaller since the intrinsic
refractive index should be smaller than that of a usual STO film due
to the oxygen vacancies [19].

The optical absorption of free electrons generated from oxygen
vacancies is not seen in the UV spectroscopy, which typically leads
to a decrease in transmittance especially in the long wavelength
region and is likely buried by the strong interference here.
However, the multiple photoluminescence (PL) emission peaks mostly
from the oxygen vacancies [20] can be clearly seen [Fig. 1(d)]. The
three PL peaks correspond to the energy intervals 2.95, 2.86 and
2.76 eV, respectively.

\begin{figure}
\includegraphics[width=3.4in]{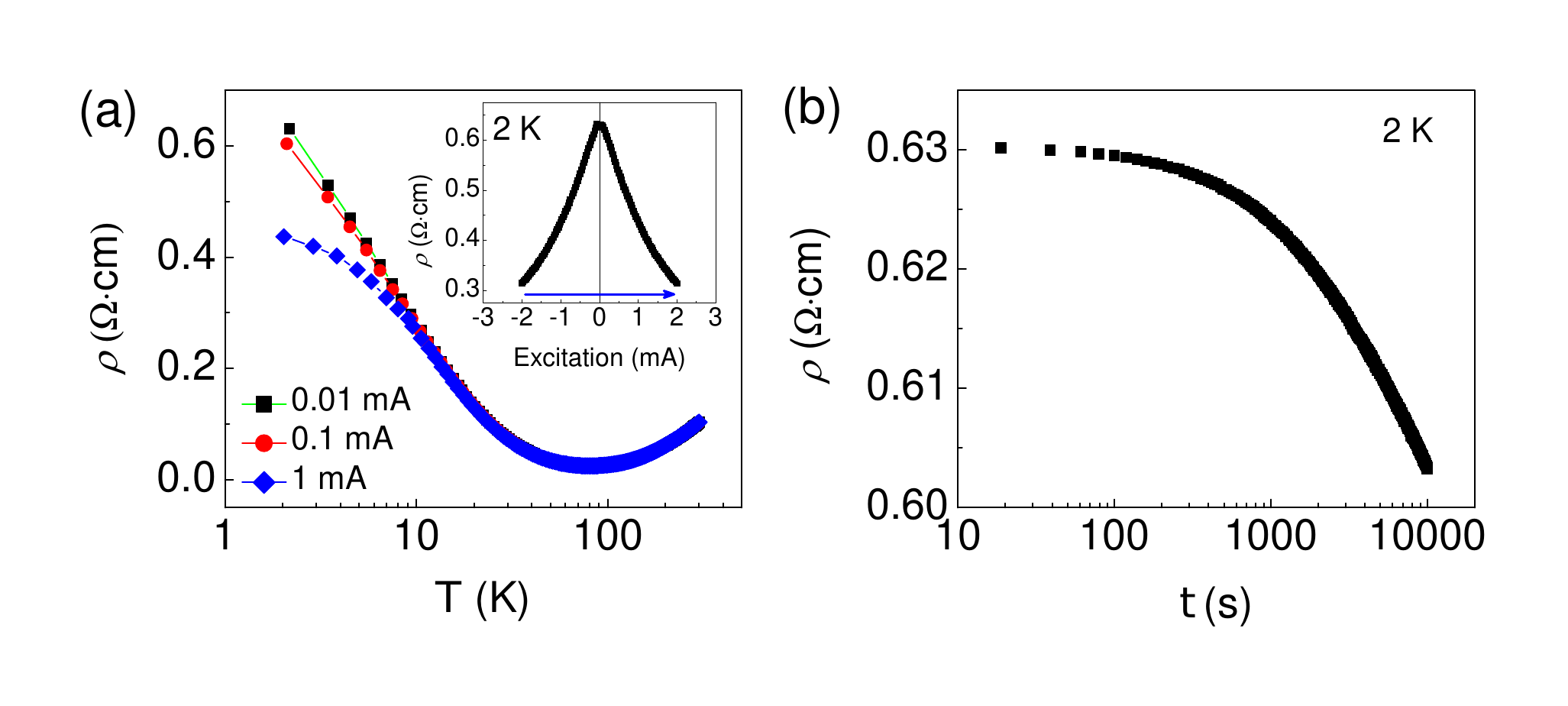}
\caption{\label{fig4}(a) $\rho-T$ curves of the oxygen deficient STO
film measured by the different currents, {\it {i.e.}}, 0.01, 0.1 and
1 mA. (Inset) The resistivity at 2 K (obtained by the $dV/dI$
measurement) versus the measuring current from -2 to 2 mA. (b) Time
dependence of the resistivity at 2 K determined by 0.01 mA for a
continues measurement up to 10 000 s.}
\end{figure}

The transport properties of the vacuum-annealed STO film were
measured by a Quantum Design PPMS machine. Aluminum wires were used
through wire bonding for contacts.  The temperature dependence of
the resistivity $(\rho-T)$ measured by 10 $\mu$A is shown on a
logarithmic scale in Fig. 2(a). There is no observable difference
between the cooling down and warming up $\rho-T$ curves. The
$\rho-T$ curves reveal an obvious metal-insulator transition at
$\sim78$ K: at higher temperatures above 78 K, the resistivity is
small and presents metallic behavior; nevertheless, the resistivity
begins to increase with decreasing temperature dramatically from 78
K and reaches nearly six times the RT resistivity at 2 K. Figure
2(b) shows that the carrier density decreases by an order of
magnitude when decreasing the temperature from 300 to 10 K.These are
well the characteristics of a carrier freeze-out phenomenon [9,21];
the density of donors (oxygen vacancies here) is low such that the
donor level is separated from the bottom of the conduction band.
Hence once the temperature decreases to some extent, most of free
electrons will shrink down to the lower donor level and get trapped.

The activation energy $\epsilon$ at high temperatures, fitted using
\begin{equation}
n\propto e^{(-\epsilon/K_B T)},
\end{equation}
is $\sim$25 meV [inset of Fig. 2(b)], which is close to the RT
thermal energy. Simply taking this value to characterize the energy
interval between the donor level and the bottom of the conduction
band, a band diagram of the oxygen deficient STO film can be
obtained as depicted in Fig. 3 on the basis of the PL emission peaks
[Fig. 1(d)].  The defect levels (close to the valence band) of STO
are consistent with [20]. The resistivity ($\sim0.1$ $\Omega$cm) and
carrier density ($1.4\times 10^{19}$ cm$^{-3}$) at RT of the STO
film are respectively $30-50$ times smaller and $\sim$30 times
larger than those of the bulk samples with the similar carrier
freeze-out phenomenon [9]. These strongly suggest that the uniform
area of the oxygen vacancies in a bulk STO is at most one-tenth of
the whole single crystal thickness. So all the previous data of the
resistivity and carrier density [6-11] related to the bulk STO
should be re-considered carefully.

The corresponding carrier mobility from 300 to 10 K [Fig. 2(a)] is
overall small and peaks around the phase transition temperature
below which the mobility decreases with decreasing temperature due
to the carrier freezing. At high temperatures the linear power law
dependence of the mobility on temperature is more obvious. The
frozen state can be re-excited by larger electric fields at low
temperatures. As shown in Fig. 4(a), the resistivity decreases with
increasing measuring current especially below 10 K. At 2 K, the
resistivity obtained from $dI/dV$ measurement clearly displays a
large negative electroresistance originated from the electrical
excitation to trapped electrons, which can be defined as
$[\rho(I)-\rho(0)]/\rho(0)$ and reaches $-50\%$ when the excitation
current is 2 mA [inset of Fig. 4(a)]. Additionally, it was found
that there is also a time dependence of the resistivity for the
frozen state. As seen in Fig. 4(b), the resistivity at 2 K relaxes
with the continuous measuring time, which suggests that the thermal
effect is playing an important role during measurement. It can be
understood by the ideas that the thermal conductivity of STO is very
low [10] over the whole temperature range from 1.4 to 100 K and
moreover after reducing there will be a significant decrease [22]
due to the scattering of phonons from oxygen vacancies. Therefore
the local temperature on the sample surface would increase as a
result of the Joule heat accumulation, eventually yielding a
decrease in resistivity under the low temperature frozen
non-metallic state. However, the thermal effect at 2 K is much
weaker than the electrical re-excitation.

The 9 T magnetic field applied perpendicular to the STO film surface
results in a negative MR although it is relatively small and can
only be seen below 5 K in the $\rho-T$ curves. The MR curve of 2 K
up to 9 T is shown in Fig. 5(a). It was found there was an asymmetry
in the first curve obtained by scanning field from -9 to 9 T. To
check whether the asymmetry was due to an improper measuring
geometry, the sample was warmed up to RT and then cooled down to 2 K
again to repeat the measurement but scanning field from 9 to -9 T.
It was found there is always an asymmetry generated from an
additional overall decrease in resistivity over the measuring time
regardless of the scanning sequence of magnetic field; this
indicates the asymmetry in the MR curves is from the thermal effect
rather than the measurement geometry. Considering deducting the
thermal effect, the shape of the MR curve at 2 K would be more close
to a letter 'M'. The positive MR under small magnetic fields can be
easily elucidated by the typical Lorentz scattering because the
magnetic field is perpendicular to the current. The positive MR is
very small because of the poor mobility of the charge carriers in
the frozen state. However, there should still be another mechanism
competing with the Lorentz scattering to account for the negative
MR. Considering that the resistance at low temperatures is
predominantly due to the trapping of carriers, so it seems plausible
to imagine that the large magnetic field can somehow help to
physically detrap the localized electrons from the trapping centers
with the assistance of an excitation electric field through the
Lorentz force.  This could be more appropriate to the specific case
here although a large number of mechanisms can result in a negative
MR.On the other hand, oxygen vacancies in the TiO$_2$ interface
layer can also enhance the tendency for ferromagnetism considerably
[23] similar to the Ti 3\emph{d} electrons from interface charge
transfer [4,24],which may also be possible for the negative MR seen
here.

The bulk STO single crystal, vacuum-annealed together with the STO
film, is metallic over the whole temperature range of $300-2$ K. The
quadratic MR at 2 K [Fig. 5(b)] is quite large up to $\sim$900\% at
5 T because of an extremely high mobility exceeding 10 000
cm$^2$V$^{-1}$s$^{-1}$ [inset of Fig. 5(b)] and does not show any
signature of a negative MR up to 5 T. This strengthens the idea that
the negative MR of the oxygen deficient STO film is closely related
to the carrier freezing state. Additionally, the bulk STO seems to
be more conductive than the STO film since there is no carrier
freeze-out in it. This evinces that the local concentration of
oxygen vacancies in the bulk STO surface is larger than that in the
STO film although they were reduced together, which strongly
suggests that there is a sharp gradient in the concentration of
oxygen vacancies in the bulk STO sample from the surface to inside.
Thus all the electrical properties related phenomena in oxygen
deficient bulk STO should be mostly just the local properties of the
near surface area.

\begin{figure}
\includegraphics[width=3.4in]{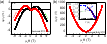}
\caption{\label{fig5} (a) Out-of-plane magnetoresistance the oxygen
deficient STO film up to 9 T and (b) the oxygen deficient STO single
crystal substrate up to 5 T which was vacuum-annealed together with
the STO film.  The temperature dependence of the carrier mobility
for the latter is shown on a logarithmic scale in the inset of (b).}
\end{figure}

Interestingly, the behavior of the carrier freeze-out observed here
is quite comparable to the spin glass behavior [25,26], for example,
the carrier freezing transition, the relaxation of the frozen
resistivity (although here is due to the thermal effect), and the
suppression of the frozen state by the external fields. From this
aspect, we can also coin this as "charge glass" to more vividly
represent its characteristics.

In summary, we argued the uniformity of oxygen vacancies in the bulk
STO single crystals by studying the high quality STO single crystal
film via various means. It was found the actual uniform thickness of
the bulk STO is around several ten times smaller than the whole
thickness and all the intriguing electrical phenomena of the oxygen
deficient bulk STO could only be the local surface properties due to
the obvious gradient in the concentrations of oxygen vacancies from
surface to inside. Moreover, we investigated the MIT observed in the
oxygen deficient STO film. The low temperature frozen state can be
remarkably re-excited by the applied electric field. The thermal
effect in oxygen deficient STO film during the electrical
measurements is pronounced due to its poor thermal conductivity
although the re-excitation of thermal effect to the low temperature
frozen state is far less effective than the electric field. It was
also found that large external magnetic fields can suppress the
carrier freezing and generate negative MR. The possible mechanism
proposed for that is the magnetic field can detrap the localized
electrons under the frozen state through Lorentz force with the help
of an electric excitation. The high similarity between the various
behaviors of the carrier freeze-out and the spin glass state enables
us to think the carrier freezing state as a kind of "charge glass"
state.

\begin{acknowledgments}
We thank the National Research Foundation (NRF) Singapore under the
Competitive Research Program (CRP) 'Tailoring Oxide Electronics by
Atomic Control' NRF2008NRF-CRP002-024, Natinoal Uinersity of
Singapore (NUS) cross-faculty grant and FRC (ARF Grant No.
R-144-000-278-112)for financial support.
\end{acknowledgments}


\end{document}